\documentclass[]{aastex631}

\usepackage{multirow}

\DeclareUnicodeCharacter{2212}{-}

\begin{document}

\title{Tip of the Red Giant Branch Distances with JWST. II. \\ I$-$band Measurements in a Sample of Hosts of 10 SN Ia Match HST Cepheids}

\author[0000-0002-8623-1082]{Siyang Li}
\affiliation{Department of Physics and Astronomy, Johns Hopkins University, Baltimore, MD 21218, USA}

\author[0000-0002-5259-2314]{Gagandeep S.~Anand}
\affiliation{Space Telescope Science Institute, 3700 San Martin Drive, Baltimore, MD 21218, USA}

\author[0000-0002-6124-1196]{Adam G.~Riess}
\affiliation{Space Telescope Science Institute, 3700 San Martin Drive, Baltimore, MD 21218, USA}
\affiliation{Department of Physics and Astronomy, Johns Hopkins University, Baltimore, MD 21218, USA}

\author{Stefano Casertano}
\affiliation{Space Telescope Science Institute, 3700 San Martin Drive, Baltimore, MD 21218, USA}

\author[0000-0001-9420-6525]{Wenlong Yuan}
\affiliation{Department of Physics and Astronomy, Johns Hopkins University, Baltimore, MD 21218, USA}

\author[0000-0003-3889-7709]{Louise Breuval}
\affiliation{Department of Physics and Astronomy, Johns Hopkins University, Baltimore, MD 21218, USA}

\author[0000-0002-1775-4859]{Lucas M.~Macri}
\affiliation{NSF's NOIRLab, 950 N Cherry Ave, Tucson, AZ 85719, USA}

\author[0000-0002-4934-5849]{Daniel Scolnic}
\affiliation{Department of Physics, Duke University, Durham, NC 27708, USA}

\author[0000-0002-1691-8217]{Rachael Beaton}
\affiliation{Space Telescope Science Institute, 3700 San Martin Drive, Baltimore, MD 21218, USA}
\affiliation{Department of Physics and Astronomy, Johns Hopkins University, Baltimore, MD 21218, USA}

\author[0000-0001-8089-4419]{Richard I.~Anderson}
\affiliation{Institute of Physics, \'Ecole Polytechnique F\'ed\'erale de Lausanne (EPFL),\\ Observatoire de Sauverny, 1290 Versoix, Switzerland}

\begin{abstract}

The Hubble Tension, a $>5\sigma$ discrepancy between direct and indirect measurements of the Hubble constant ($H_0$), has persisted for a decade and motivated intense scrutiny of the paths used to infer $H_0$. Comparing independently-derived distances for a set of galaxies with different standard candles, such as the tip of the red giant branch (TRGB) and Cepheid variables, can test for systematics in the middle rung of the distance ladder. The $I$ band is the preferred filter for measuring the TRGB due to constancy with color, a result of low sensitivity to population differences in age and metallicity supported by stellar models. We use \emph{James Webb Space Telescope (JWST)} observations with the maser host NGC$\,$4258 as our geometric anchor to measure $I$-band (\emph{F090W} vs \emph{F090W}$-$\emph{F150W}) TRGB distances to 8 hosts of 10 Type Ia supernovae (SNe Ia) within 28 Mpc: NGC$\,$1448, NGC$\,$1559, NGC$\,$2525, NGC$\,$3370, NGC$\,$3447, NGC$\,$5584, NGC$\,$5643, and NGC$\,$5861. We compare these with \emph{Hubble Space Telescope (HST)} Cepheid-based relative distance moduli for the same galaxies and anchor. We find no evidence of a difference between their weighted means, $0.01$ $\pm$ 0.04 (stat) $\pm$ $0.04$ (sys) mag. We produce fourteen variants of the TRGB analysis, altering the smoothing level and color range used to measure the tips to explore their impact.  For some hosts, this changes the identification of the strongest peak, but this causes little change to the sample mean difference producing a full range of 0.00 to 0.02 mag, all consistent at 1$\sigma$ with no difference. The result matches past comparisons of $I$-band TRGB and Cepheids when both use \emph{HST}.  SNe and anchor samples observed with \emph{JWST} are too small to yield a measure of $H_0$ that is competitive with the \emph{HST} sample of 42 SNe Ia and 4 anchors; however, they already provide a vital systematic crosscheck to \emph{HST} measurements of the distance ladder.
\end{abstract}


\section{Introduction} \label{sec:intro}

The `Hubble Tension' refers to a 5$-$6 $\sigma$ discrepancy between direct measurements and cosmological inference of the present expansion rate of the Universe, parameterized as the Hubble constant ($H_0$). Currently, the ``gold standard'' method to directly measure $H_0$ utilizes a three-rung ladder to obtain distances to galaxies in the Hubble flow. This method uses geometric distances derived from parallaxes (such as those from \citealt{Gaia_2021A&A...649A...1G}), water masers \citep{Reid_2019ApJ...886L..27R}, and detached eclipsing binaries \citep{Pietrzynski_2019Natur.567..200P, Gracyzk_2020ApJ...904...13G} to calibrate the luminosities of primary distance indicators such as Cepheid variables (first rung). This calibration is in turn used to measure distances to galaxies that host both Cepheids and Type Ia supernovae to calibrate the luminosities of the latter (second rung), which are then finally used to measure distances to galaxies in the Hubble flow to measure $H_0$ (third rung). The Supernovae, $H_0$, for the Equation of State (SH0ES) team, for instance, has used this method to measure $H_0$ = 73.17 $\pm$ 0.86 km s$^{−1}$ Mpc$^{−1}$ \citep{Riess_2022ApJ...934L...7R,Breuval_2024arXiv240408038B}. The cosmological inference uses the cosmic microwave background together with a cosmological model such as $\Lambda$CDM to infer the present expansion rate. \citet{Planck_2020A&A...641A...6P} used this approach to derive $H_0=$ 67.36 $\pm$ 0.54 km s$^{−1}$ Mpc$^{−1}$, a $>5\sigma$ difference from the value obtained by \cite{Breuval_2024arXiv240408038B}. Several independent teams and measurements using a wide range of distance indicators have also yielded measurements in tension with those from {\it Planck}. No precise direct measurement of $H_0$ has yielded a value that is lower than that from {\it Planck} (see compilations in \citealt{Di_valentino_2021CQGra..38o3001D, Verde_2023arXiv231113305V}). This tension suggests a possible need to revise the standard $\Lambda$CDM model, though efforts continue to search for unidentified systematics in the measurements at any level\footnote{For an extensive list of studies that have studied systematics related to the tension, we refer the reader to \url{https://djbrout.github.io/SH0ESrefs.html}.}. 

Independent distance measurements to the same galaxies using multiple standard candles offers one of the best ways to ensure systematics are well accounted for and understood. The tip of the red giant branch (TRGB) refers to the maximum luminosity reached by first-ascent red giant stars before transitioning onto the horizontal branch, and has been used as a standard candle to measure distances \citep[e.g.][]{
Anand_Maffei_2019ApJ...872L...4A,
Anand_Local_Group_2019ApJ...880...52A, 
Beaton_M101_2019ApJ...885..141B, 
Hoyt_N5643_N1404_2021ApJ...915...34H, 
Lee_WLM_2021ApJ...907..112L, Anand_EDD_2021AJ....162...80A, 
Anand_PHANGS_2021MNRAS.501.3621A, 
Oakes_Fornax_2022ApJ...929..116O, 
Tran_Sculptor_2022ApJ...935...34T,
Lee_M33_2022ApJ...933..201L, Madore_Ceph_TRGB_Metallicity_2024ApJ...961..166M, Anand_2024ApJ...966...89A, Anand_Fornax_2024arXiv240503743A}, and the Hubble constant \cite[e.g.][see also review in \citealt{Hubble_Tension_Book_2024}]{Freedman_2019ApJ...882...34F, Kim_TRGB_Virgo_H0_2020ApJ...905..104K, Freedman_2021ApJ...919...16F, Blakeslee_2021ApJ...911...65B,Dhawan_TRGB_H0_2022ApJ...934..185D,  Anand_2022ApJ...932...15A, Dhawan_BayeSN_H0_2023MNRAS.524..235D, Scolnic_2023ApJ...954L..31S}. The TRGB can thus be used as a means to crosscheck systematics of other standard candles, such as Cepheid variables.

The TRGB is commonly measured in the \emph{I} band, or the \emph{Hubble Space Telescope (HST)} \emph{F814W} and \emph{James Webb Space Telescope (JWST)} \emph{F090W} equivalents, where the magnitude of the TRGB exhibits the smallest metallicity (and hence color) dependence \citep{Rizzi_2007ApJ...661..815R, Jang_2017ApJ...835...28J}. While it is possible to measure the TRGB in the NIR to take advantage of brighter magnitudes, doing so requires careful calibration of the TRGB's color dependence \citep{Bellazzini_2004A&A...424..199B,Dalcanton_2012ApJS..198....6D,Wu_2014AJ....148....7W,Madore_IC1613_NIR_TRGB_2018ApJ...858...11M, Hoyt_LMC_NIR_TRGB_2018ApJ...858...12H, McQuinn_2019ApJ...880...63M, Newmann_HST_2024ApJ...966..175N, Newmann_JWST_2024arXiv240603532N, Hoyt_2024arXiv240707309H} where it rises to $\sim$ one magnitude per magnitude of measured color, introducing additional systematics. In this study, we focus on measuring the TRGB in the \emph{JWST} \emph{F090W} filter, which is most similar to \emph{I} and \emph{F814W} used with \emph{HST}.

The launch of \emph{JWST} has opened up the possibility of performing extensive crosschecks of Cepheids, TRGB, and other standard candles in unprecedented resolution. In particular, the dual-module configuration of the \emph{JWST} NIRCam instrument allows for at least three standard candles (Cepheids, TRGB, and Carbon stars) to be simultaneously observed with one set of observations $-$ see for instance, \emph{JWST} programs GO-1685 \citep{Riess_2021jwst.prop.1685R}, GO-1995 \citep{Freedman_2021jwst.prop.1995F}, and GO-2875 \citep{Riess_2023jwst.prop.2875R}, among others.
In a previous study, \cite{Anand_2024ApJ...966...89A} (Paper I), we measured the TRGB with \emph{JWST} in the maser host NGC$\,$4258, which serves as one of four geometric calibrators of the distance ladder, and presented measurements in 2 SN Ia hosts (NGC$\,$1559 and NGC$\,$5584). The derived luminosity of the TRGB in \emph{JWST} \emph{F090W} in Paper 1 matched that independently determined by \cite{Newmann_HST_2024ApJ...966..175N}, also in \emph{F090W}, to $\pm$ 0.01 mag, indicating a reliable foundation for distance measurements with JWST in SN Ia hosts.
Here we study an expanded sample using \emph{JWST} observations of 8 hosts of 10 SNe~Ia from Cycle 1 program GO-1685 \citep{Riess_2021jwst.prop.1685R} and Cycle 2 GO-2785 \citep{Riess_2023jwst.prop.2875R} to measure TRGB-based distances to these galaxies and test their consistency with \emph{HST} Cepheid-based distances to those same galaxies.

\section{Data} \label{sec:Data}

\begin{figure}
\epsscale{1.1}
\plotone{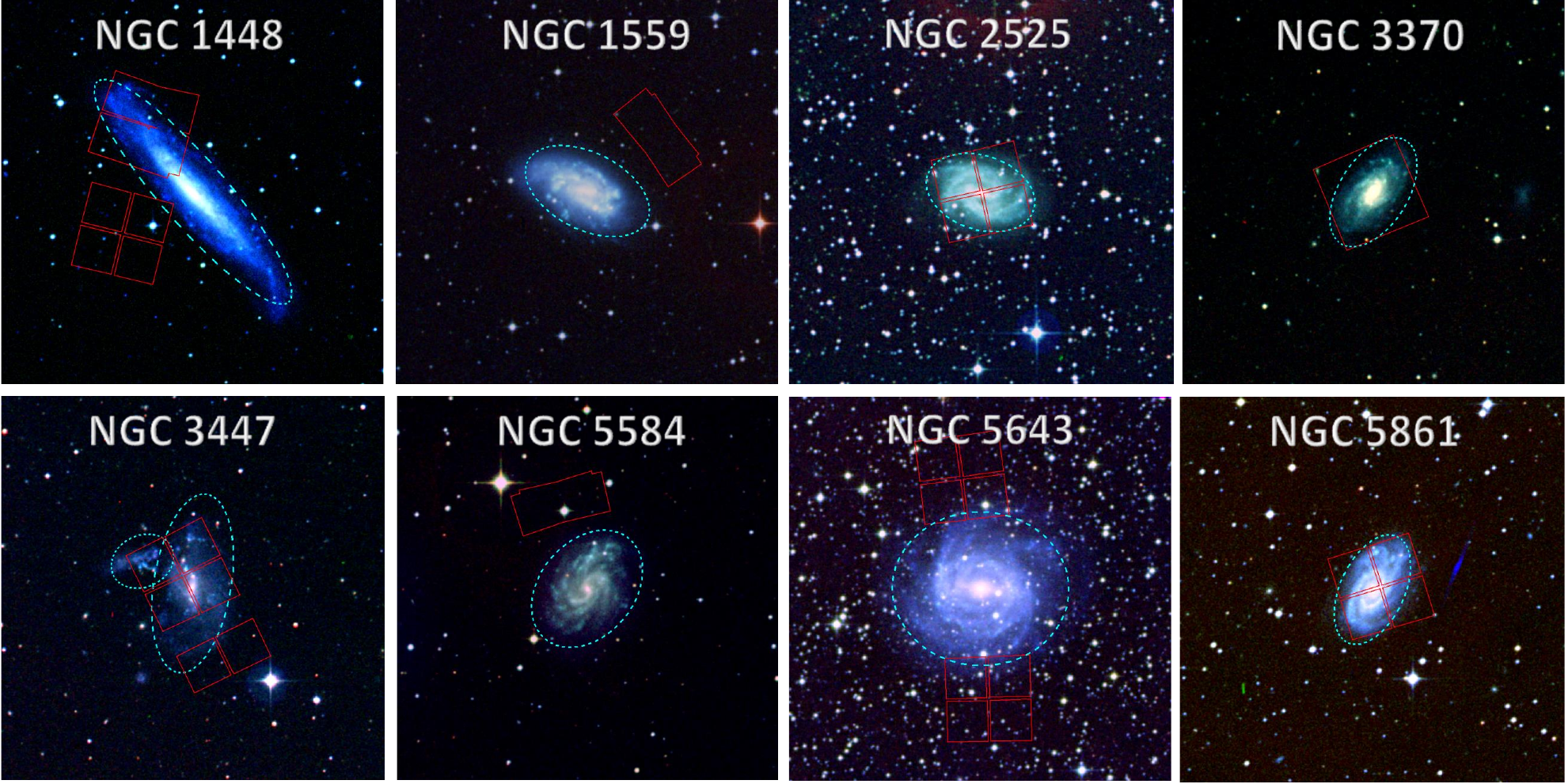}
\caption{Footprints of the portions of the \emph{JWST}/NIRCam visits used for our analysis (see the main text for details of our spatial selections), overlaid on $10\arcmin \times 10\arcmin$ images from the Digitized Sky Survey. The cyan dashed lines show $D_{25}$ (except for NGC$\,$3447, which is an interacting system), and we remove any stars within this region from our catalogs prior to our measurements. }
\label{fig:footprints}
\end{figure}

We retrieve \emph{JWST} \emph{F090W} and \emph{F150W} observations of NGC$\,$1448, NGC$\,$1559, NGC$\,$5584, and NGC$\,$5643 from GO-1685 \citep{Riess_2021jwst.prop.1685R} and NGC$\,$2525, NGC$\,$3370, NGC$\,$3447, and NGC$\,$5861 from GO-2875 \citep{Riess_2023jwst.prop.2875R} using the Barbara A.~Mikulski Archive for Space Telescopes (MAST)\footnote{\url{https://mast.stsci.edu/portal/Mashup/Clients/Mast/Portal.html}}. We list observation details in Table \ref{tab:Observations_Table} and show galaxy footprints in Fig.~\ref{fig:footprints}. For NGC$\,$2525, we found that the images corresponding to the fourth (and final) dither for each filter were noticeably blurred. Upon further investigation, we found that the guide star had wandered onto some bad pixels in the Fine Guidance Sensor. We thus excluded these dithers from our analysis and generated our own stage 3 image from only the first three dithers for use of mutual image alignment of the underlying stage 2 images with DOLPHOT. 

%%%%%%%%%%%%%%%%%%%%%%%%%%%%%%%%%%%%%%
%%%%%%%%%%%%%%%%%%%%%%%%%%%%%%%%%%%%%%

%%%%%%%%%%%%%%%%%%%%%%%%%%%%%%%%%%%%%%
\begin{deluxetable*}{cccccccc}
\label{tab:Observations_Table}
\tablehead{\colhead{Galaxy} & \colhead{Program} & \colhead{Observation Number} & \colhead{Observation Date} & \colhead{Filters} & \colhead{Exposure Time [s]} & \colhead{\texttt{pmap} Version}}
\startdata
NGC$\,$1448 & GO-1685 & 13 & 2023-08-02 & \emph{F090W}/\emph{F277W} & 418.7 x 4 & 1215 \\
NGC$\,$1448 & GO-1685 & 13 & 2023-08-02 & \emph{F150W}/\emph{F277W} & 526.1 x 4 & " \\
NGC$\,$1448 & GO-1685 & 14 & 2023-18-02 & \emph{F090W}/\emph{F277W} & 418.7 x 4 & " \\
NGC$\,$1448 & GO-1685 & 14 & 2023-18-02 & \emph{F150W}/\emph{F277W} & 526.1 x 4 & " \\
NGC$\,$1559 & GO-1685 & 1 & 2023-06-30 & \emph{F090W}/\emph{F277W} & 418.7 x 4 & " \\
NGC$\,$1559 & GO-1685 & 1 & 2023-06-30 & \emph{F150W}/\emph{F277W} & 526.1 x 4 & " \\
NGC$\,$1559 & GO-1685 & 2 & 2023-07-15 & \emph{F090W}/\emph{F277W} & 418.7 x 4 & " \\
NGC$\,$1559 & GO-1685 & 2 & 2023-07-15 & \emph{F150W}/\emph{F277W} & 526.1 x 4 & " \\
NGC$\,$5584 & GO-1685 & 9 & 2023-01-30 & \emph{F090W}/\emph{F277W} & 418.7 x 4 & "  \\
NGC$\,$5584 & GO-1685 & 9 & 2023-01-30 & \emph{F150W}/\emph{F277W} & 526.1 x 4 & "  \\
NGC$\,$5584 & GO-1685 & 10 & 2023-02-21 & \emph{F090W}/\emph{F277W} & 418.7 x 4 & "  \\
NGC$\,$5584 & GO-1685 & 10 & 2023-02-21 & \emph{F150W}/\emph{F277W} & 526.1 x 4 & "  \\
NGC$\,$5643 & GO-1685 & 11 & 2023-07-07 & \emph{F090W}/\emph{F277W} & 311.4 x 4 & "  \\
NGC$\,$5643 & GO-1685 & 11 & 2023-07-07 & \emph{F150W}/\emph{F277W} & 418.7 x 4 & "  \\
NGC$\,$5643 & GO-1685 & 12 & 2023-07-22 & \emph{F090W}/\emph{F277W} & 311.4 x 4 & "  \\
NGC$\,$5643 & GO-1685 & 12 & 2023-07-22 & \emph{F150W}/\emph{F277W} & 418.7 x 4 & "  \\
NGC$\,$2525 & GO-2875 & 1 & 2023-04-23 & \emph{F090W}/\emph{F277W} & 472.4 x 3 & 1225 \\
NGC$\,$2525 & GO-2875 & 1 & 2023-04-23 & \emph{F150W}/\emph{F277W} & 526.1 x 3 & " \\
NGC$\,$3370 & GO-2875 & 2 & 2023-06-28 & \emph{F090W}/\emph{F277W} & 472.4 x 4 & 1234 \\
NGC$\,$3370 & GO-2875 & 2 & 2023-06-28 & \emph{F150W}/\emph{F277W} & 526.1 x 4 & " \\
NGC$\,$3447 & GO-2875 & 12 & 2023-05-24 & \emph{F090W}/\emph{F277W} & 472.4 x 4 & "  \\
NGC$\,$3447 & GO-2875 & 12 & 2023-05-24 & \emph{F150W}/\emph{F277W} & 622.7 x 4 & "  \\
NGC$\,$5861 & GO-2875 & 14 & 2024-07-30 & \emph{F090W}/\emph{F277W} & 472.4 x 4 & "  \\
NGC$\,$5861 & GO-2875 & 14 & 2024-07-30 & \emph{F150W}/\emph{F277W} & 622.7 x 4 & "  \\
\enddata
\caption{Summary table for the \emph{JWST} NIRCam observations of NGC$\,$1448, NGC$\,$1559, NGC$\,$2525, NGC$\,$3370, NGC$\,$3447, NGC$\,$5584, NGC$\,$5643, and NGC $\,$5861 used in this study. Columns from left to right are: galaxy name, program number, observation number, observation date, filters, exposure time per dither, and context (\texttt{pmap}) file version.}
\end{deluxetable*}
%%%%%%%%%%%%%%%%%%%%%%%%%%%%%%%%%%%%%%
\newpage
\section{Photometry} \label{sec:Photometry}

We follow the general procedures outlined by previous works from our team for photometric reductions \citep{Riess_2024ApJ...962L..17R,Anand_2024ApJ...966...89A}. In detail, we perform PSF photometry\footnote{The photometry for this study is available publicly upon publication in a Zenodo repository, DOI: 10.5281/zenodo.13131990.} using the DOLPHOT\footnote{\url{http://americano.dolphinsim.com/dolphot/}} software package \citep{Dolphin_2000PASP..112.1383D, Dolphin_2016ascl.soft08013D}, together with the latest (February 2024) version of the \emph{JWST}/NIRCam module \citep{Weisz_2024ApJS..271...47W}. We use the stage 3 \emph{F150W} \texttt{i2d} files as reference frames and perform photometry directly on the stage 2 \texttt{cal} images.  We use the Vega-Vega system to remain consistent with earlier versions of DOLPHOT's \emph{JWST} module, which gives a $\sim$0.04~mag offset (in \emph{F090W}) from the Vega-Sirius system that is currently default in DOLPHOT NIRCam module. We use the ``-etctime" option, which allows the package to use (\texttt{TMEASURE}) for S/N calculations. We apply the following DOLPHOT quality cuts, similar to those from \cite{Weisz_2024ApJS..271...47W}: \texttt{crowd} $<$ 0.5 (in both bands), \texttt{sharp$^2$} $<$ 0.01 (in both bands), and \texttt{type} $\le$ 2. In addition to these quality cuts, we also apply S/N cuts in both bands for each target. For NGC$\,$1448, NGC$\,$1559, NGC$\,$3447, NGC$\,$5584, and NGC$\,$5643, we adopt a minimum threshold of \texttt{S/N}=5, whereas we choose a value of \texttt{S/N}=3 for NGC$\,$2525, NGC$\,$3370, and NGC$\,$5861 (due to the decreased relative depth of their photometry). 

We also correct for foreground extinction using \emph{E(B-V)} from \cite{Schlafly_2011ApJ...737..103S} and galaxy coordinates, both retrieved using the the NASA/IPAC Extragalactic Distance Database (NED)\footnote{\url{https://ned.ipac.caltech.edu/}}and listed in Table \ref{tab:Distances} and the TRGB color selection has been adjusted to account for the reddening correction. We use the \cite{Fitzpatrick_1999PASP..111...63F} reddening law with $R_v$ = 3.1 and $A_{\lambda}/$\emph{E(B-V)} = 1.4156 and 0.6021 for \emph{F090W} and \emph{F150W}, respectively \citep{Anand_2024ApJ...966...89A}. We adopt the extinction used in \cite{Anand_2022ApJ...932...15A} of \emph{E(B-V)} = 0.161~mag for NGC 5643, which is located at a low galactic latitude (where the dust maps are less certain), and where the adopted value was re-estimated using the displacement of the zero-age main sequence following \cite{Rizzi_2017ApJ...835...78R}.

\section{Measurement} \label{sec:Measurement}

Measuring the TRGB magnitude involves locating the discontinuity or inflection point of the luminosity function consisting of red giant and asymptotic giant branch stars. This can be done many ways; for instance, using a Sobel filter \citep{Lee_1993ApJ...417..553L, Hatt_2017ApJ...845..146H}, or model-based methods, with either least-squares fitters \citep{Wu_2014AJ....148....7W, Crnojevic_2019ApJ...872...80C} or maximum-likelihood estimation \citep{Makarov_2006AJ....132.2729M, Li_2022ApJ...939...96L, Li_2023ApJ...950...83L}. In this study, we measure the TRGB using a Sobel filter using the measurement routine without spatial clipping (which we instead perform using elliptical annuli) from the Comparative Analysis of TRGBs (CATs) team \citep{Wu_2023ApJ...954...87W, Scolnic_2023ApJ...954L..31S, Li_CATs_2023ApJ...956...32L}, which is publicly available on GitHub\footnote{\url{https://github.com/JiaxiWu1018/Unsupervised-TRGB}}. The CATs team had optimized their measurement parameters (such as smoothing value, color cuts, etc.) by minimizing field-to-field dispersion in the TRGB measurement across multiple fields. At the moment, there are a limited number of \emph{JWST} fields in each host galaxy, which prevents us from performing a similar optimization. For this study, we adopt a smoothing value of 0.05~mag and unweighted, or `simple' weighting as named in the CATs algorithm (i.e. no further weighting of the output of the edge-detection measurement). We note that without the CATs optimizations in place, this methodology is equivalent to the TRGB measurements adopted by the Carnegie-Chicago Hubble Program \citep[CCHP;][]{Beaton_CCHP_Overview_2016ApJ...832..210B, Freedman_2019ApJ...882...34F}, but without the additional weighting of the Sobel filter output (which has been shown to cause biases in the outputs; \citealt{Anderson_2024ApJ...963L..43A, Anand_2024ApJ...966...89A}). We explore the effects of varying the level of smoothing of the luminosity function and the color range used to measure the luminosity function, 14 combinations in all, in the next section. TRGB uncertainties are estimated using bootstrap resampling of the luminosity function with 5,000 samples and the standard deviation of the bootstrap distribution. These distributions all follow Gaussianity with means that lie within 1 $\sigma$ from the measured TRGB. For NGC 5584, using the smaller smoothing value of $s$=0.05~mag produces a Sobel response that shows two peaks of nearly the same height. Following \cite{Anand_2024ApJ...966...89A}, if there are two Sobel peaks that are of equivalent height ($\leq$ 3$\%$), we select the brighter peak. The bootstrap distribution for NGC 5584 exhibits two Gaussians corresponding to the two Sobel peaks in the baseline result. We incorporate this spread to account for uncertainty in peak selection for NGC 5584.

Before measuring the TRGB, we perform spatial selections to limit contamination from younger stellar populations in these star-forming galaxies. The NIRCam regions used to measure the TRGB are highlighted in Fig.~\ref{fig:footprints} $-$ in all cases (except NGC$\,$3447, described below), only regions exterior to the $B=25$~mag/sq arcsec isophote ($D_{25}$; \citealt{RC3_1991rc3..book.....D}) are selected \citep{Anand_2022ApJ...932...15A, Anand_2024ApJ...966...89A, Hoyt_2024arXiv240707309H}. For NGC$\,$1559 and NGC$\,$5584, the parallel NIRCam module contains a suitable sampling of stars, although only half of the parallel module is used (the far half of each contain too few stars to perform precise empirical PSF adjustments and aperture corrections). The same is nearly true for NGC$\,$5643, though we remove a small portion of each parallel module that is interior to $D_{25}$. In NGC$\,$1448, NGC$\,$2525, NGC$\,$3370, and NGC$\,$5861, a sizeable portion of the main module is outside of $D_{25}$, which we include to increase the sampling of stars and for NGC$\,$2525, NGC$\,$3370 and NGC$\,$5861, the parallel modules contain too few stars to perform photometry on. Lastly, NGC$\,$3447 is an actively interacting system, often described as two targets, N3447A and B. Here, we use two ellipses centered at RA, Dec = (10:53:23.998, +16:46:26.78) and (10:53:29:3934, +16:47:02.294), major axes of 140 and 50 arcsec, minor axes of 60 and 40 arcsec, and position angles of 80 and 115 degrees, respectively, following \cite{Mazzei_2018A&A...610A...8M}. 

\begin{figure}
\epsscale{1.1}
\plotone{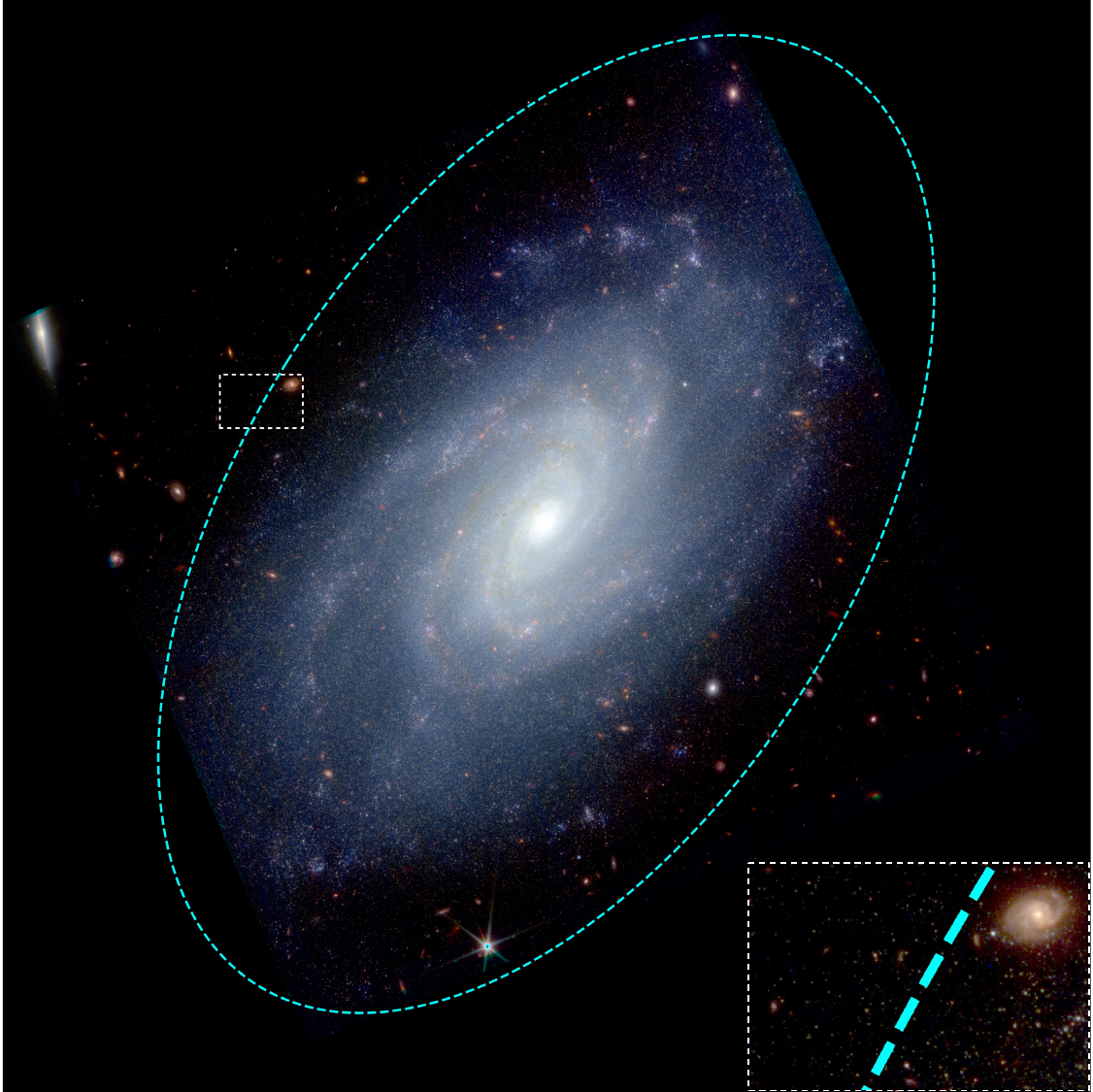}
\caption{Color image of NGC$\,$3370 generated from our $F090W$+$F150W$+$F277W$ \emph{JWST}/NIRCam imaging. $D_{25}$ ($B=25$~mag/sq arcsec isophote) is shown as the cyan dashed line $-$ only stars outside of this region are included in our TRGB analysis. A dashed white inset shows a zoom-in on a region near the selected isophote (the color inset image is displayed with a tighter dynamic range to more clearly show stars in the halo).}
\label{fig:n3370_color}
\end{figure}

A color image of NGC$\,$3370 is shown in Fig.~\ref{fig:n3370_color}, along with the same $D_{25}$ selection region adopted for the spatial cuts. It can be seen that the outer regions of the image used for our analysis are of low stellar density, where stars are easily separated. While in some cases the TRGB is only $\sim$1 magnitude above the photometric completeness limit, we note that our spatial selection criteria as described above limits the effects of photometric bias, which would otherwise skew our edge-detection measurements in higher surface-brightness regions. Even for galaxies where the TRGB is located closest to the photometric limit, we do not measure any noticeable photometric bias in our selected halo regions–– for example, the TRGB in NGC~3370 lies near a photometric completeness of $\sim$65$\%$, but the measured photometric bias remains negligible ($<$0.01~mag).

%%%%%%%%%%%%%%%%%%%%%%%%%%%%%%%%%%%%%%
%%%%%%%%%%%%%%%%%%%%%%%%%%%%%%%%%%%%%%

The TRGB in \emph{F090W} is expected to have minimal variation ($<$0.02~mag) with color over a range of 1.15 $<$ \emph{F090W} $-$ \emph{F150W} $<$ 1.75~mag \citep{Anand_2024ApJ...966...89A}. Here we adopt a color selection of 1.30 $<$ \emph{F090W} $-$ \emph{F150W} $<$ 1.75~mag for our baseline result (see Fig.~2 in \citealt{Anand_2024ApJ...966...89A}), where the tighter range aims to limit the effects of a limited amount of supergiants which remain in our CMDs.

%%%%%%%%%%%%%%%%%%%%%%%%%%%%%%%%%%%%%%
\begin{figure*}[ht!]
  \centering
  \begin{tabular}{cc}
    \includegraphics[width=0.37\linewidth]{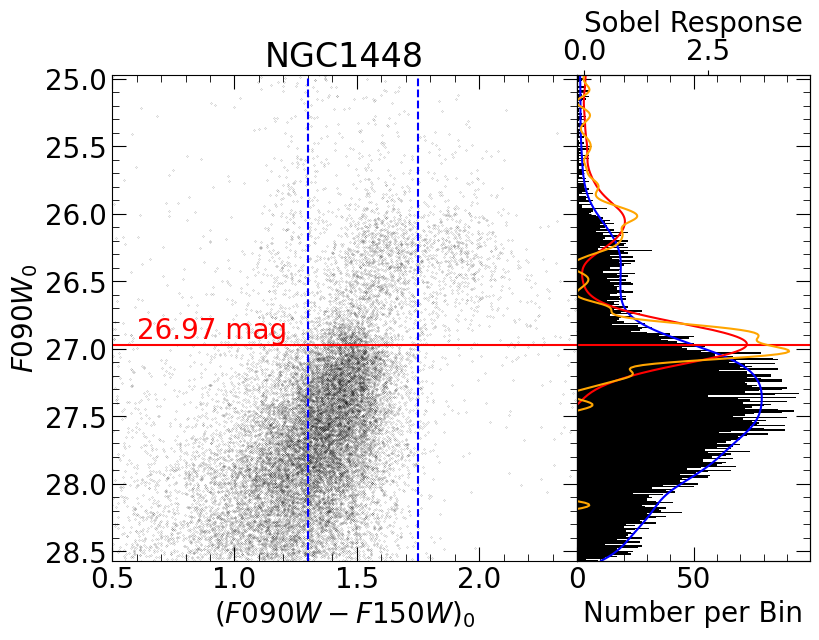} &
    \includegraphics[width=0.37\linewidth]{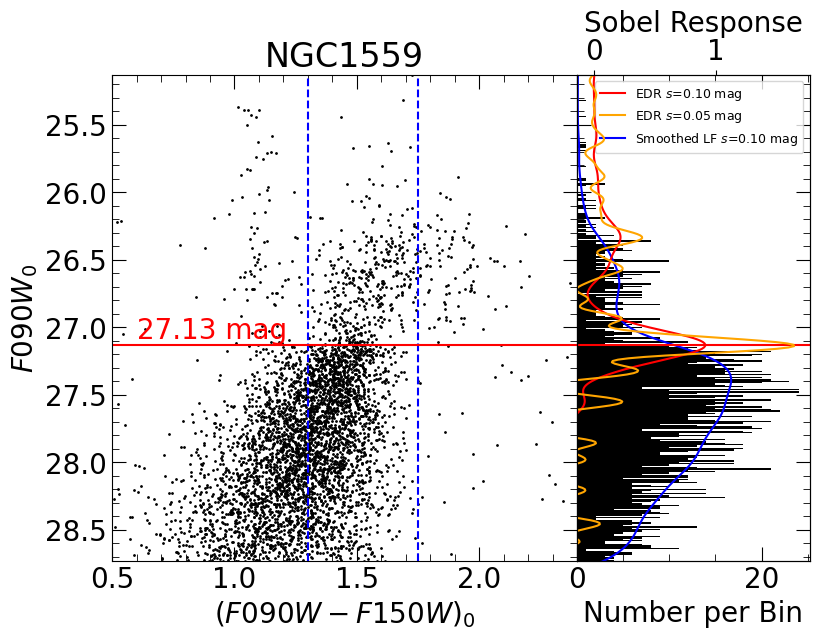} \\
    \includegraphics[width=0.37\linewidth]{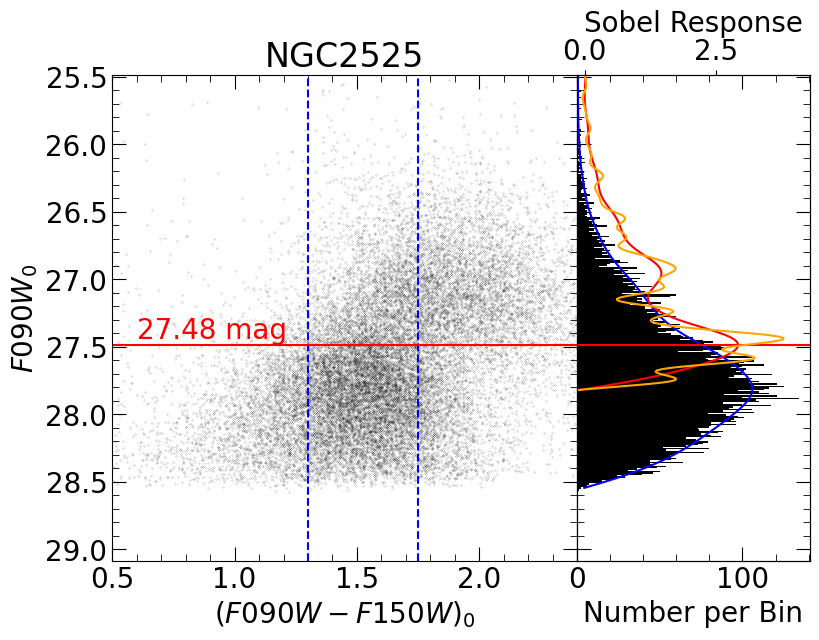} &
    \includegraphics[width=0.37\linewidth]{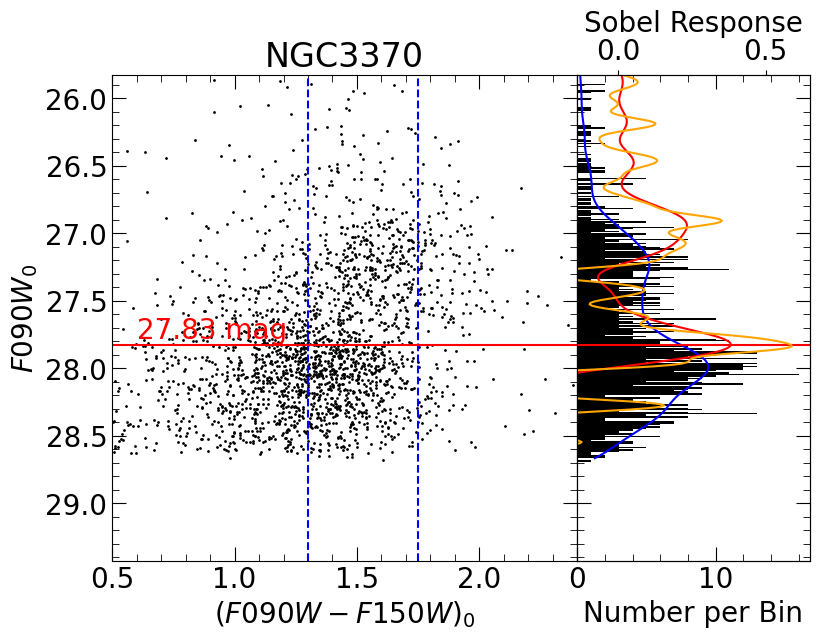} \\
    \includegraphics[width=0.37\linewidth]{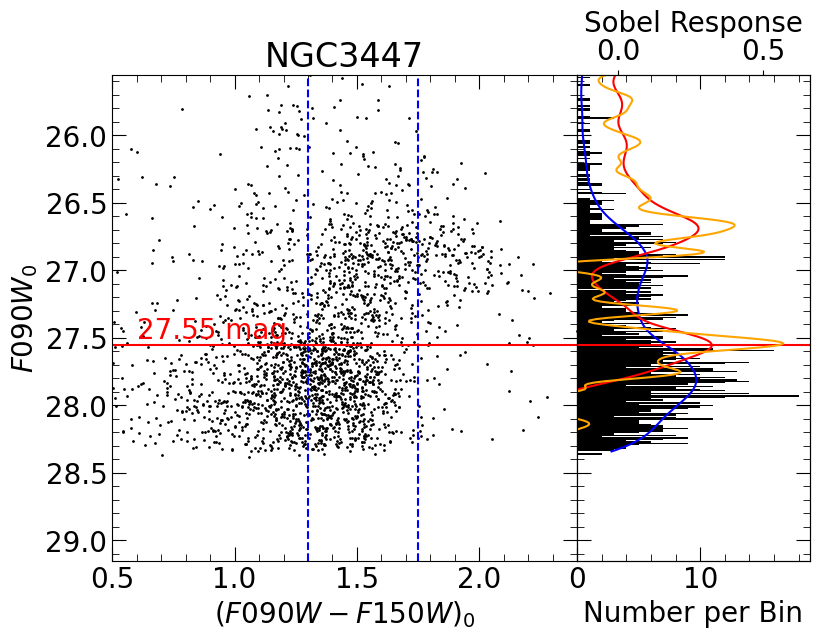} &
    \includegraphics[width=0.37\linewidth]{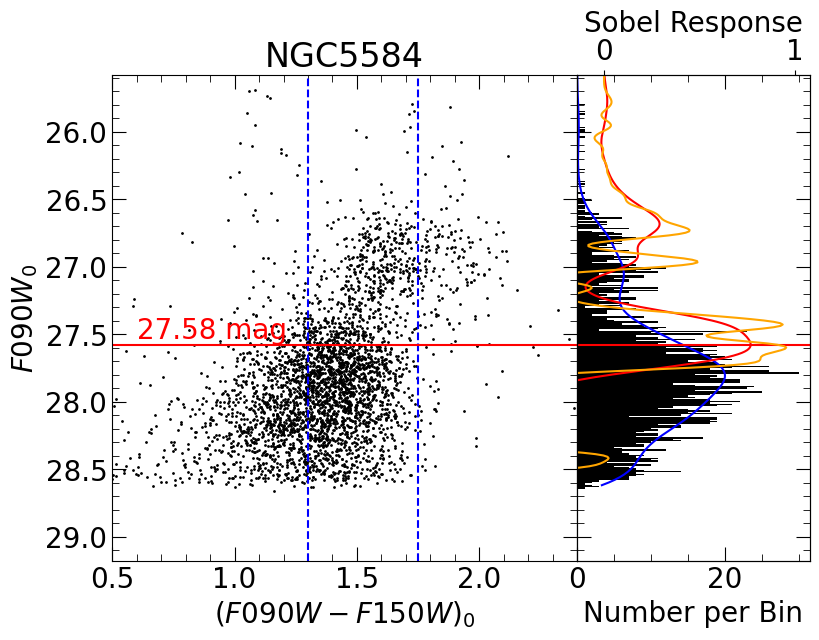} \\
    \includegraphics[width=0.37\linewidth]{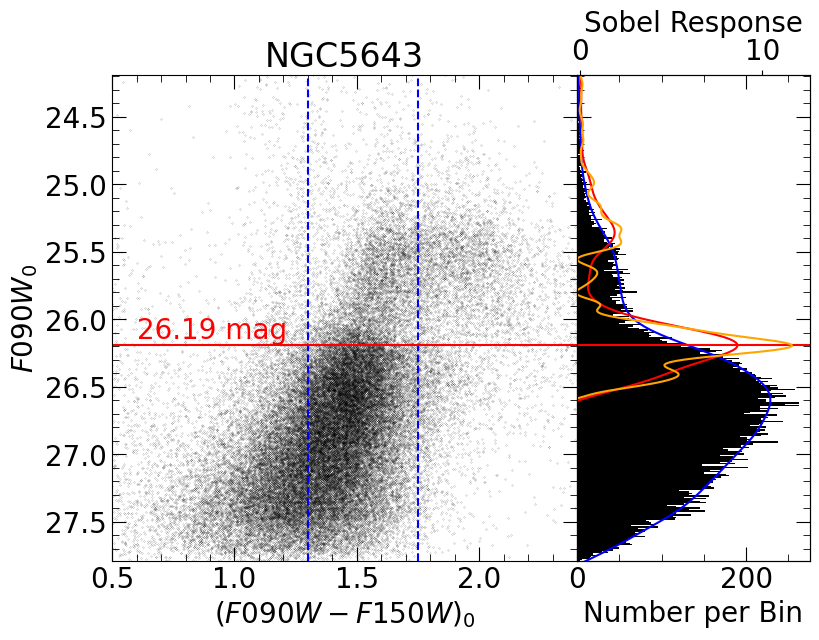} &
    \includegraphics[width=0.37\linewidth]{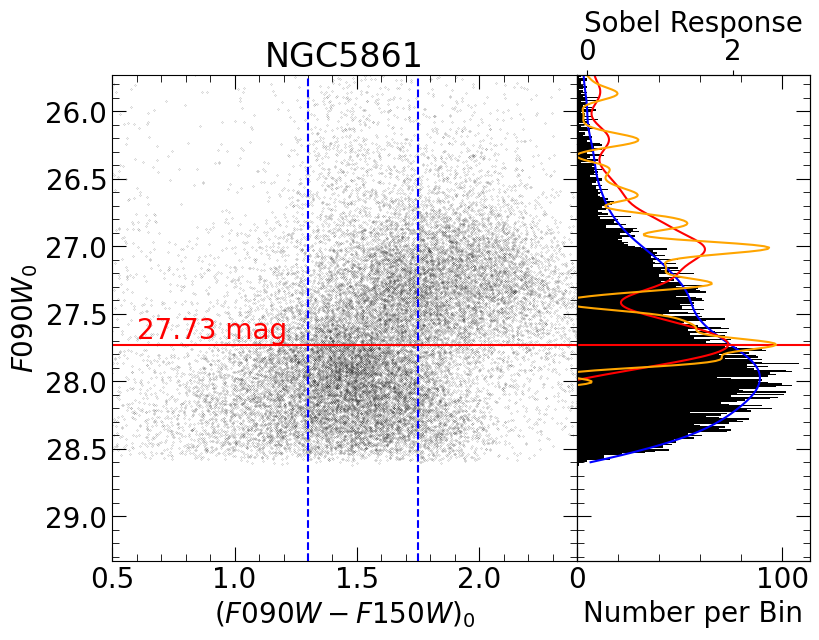} \\
    \multicolumn{2}{c}{\includegraphics[width=0.7\linewidth]{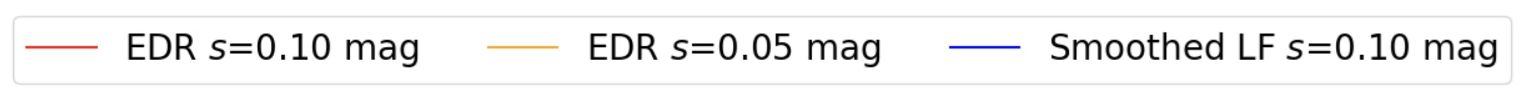}}
  \end{tabular}
  \caption{Color magnitude diagrams and luminosity functions used for the TRGB measurements in this study. The vertical blue dashed lines show the color cuts of 1.30 $<$ \emph{F090W $-$ F150W} $<$ 1.75~mag, and the horizontal red line and value show the location of the measured TRGB using $s$=0.10~mag. The blue curve shows the smoothed ($s$=0.10~mag) luminosity function (LF). The orange and red curves show the response of the Sobel filter applied on luminosity functions binned in 0.01~mag intervals (black) using smoothing values of 0.05 and 0.10~mag, respectively. EDR in the legend refers to the edge detector response (EDR) using a Sobel filter.}
  \label{fig:CMDs}
\end{figure*}
%%%%%%%%%%%%%%%%%%%%%%%%%%%%%%%%%%%%%%

% \clearpage

\section{Distances} \label{sec:Distances}

We use the TRGB measurements shown in Fig.~\ref{fig:CMDs} to measure distances to the host galaxies. We adopt a TRGB zero-point of $M^{F090W}_{TRGB}$ = $-4.372 \pm 0.033$ (stat) $\pm~0.045$ (sys)~mag from \cite{Anand_2024ApJ...966...89A} $-$ based on $m_{TRGB} = 25.045$~mag, which is the mean of the two simple Sobel measurements using a D $>$ $D_{25}$ and outer spatial selections (either being within $\pm$ 0.005~mag of this mean). This calibration uses the geometric maser distance to NGC$\,$4258 \citep{Reid_2019ApJ...886L..27R}, to calibrate the TRGB in the \emph{JWST F090W} system and is stable for smoothing values $s=0.04$~mag to $s=0.12$~mag due to the absence of weighting (see Fig. 6 in \citealt{Anand_2024ApJ...966...89A}). This absolute calibration is also highly consistent with the independently derived zero-point of $M^{F090W}_{TRGB}$ = $-4.36 \pm  0.025$ (stat) $\pm$ 0.043 (sys)~mag from \cite{Newmann_JWST_2024arXiv240603532N}, after adjusting their value from the Sirius$-$Vega system to the Vega$-$Vega zeropoint system (M. Newman, priv. communication).  We list and plot the distances to the host galaxies, compared to Cepheid-based distances from \cite{Riess_2022ApJ...934L...7R}. We compare to fit variant 10 from \citep{Riess_2022ApJ...934L...7R} for consistency in anchor selection; the distances produced by fit variant 10 are on average greater than those using all anchors (baseline) by $\sim$0.015 mag, consistent with their differences in $H_0$ of 72.5 km s$^{−1}$ Mpc$^{−1}$ for NGC$\,$4258 and 73.0 km s$^{−1}$ Mpc$^{−1}$ using all anchors.

We find a weighted mean difference between the TRGB (baseline results with $1.30 < F090W-F150W<1.75$~mag and smoothing of $s=0.05$~mag) and {\it HST} Cepheid distances of 0.01 $\pm$ 0.04 (stat) $\pm$ 0.04 (sys)~mag, which includes the systematic error due to measurement choice listed in Table \ref{tab:Distances}, between the TRGB distance moduli measured here and Cepheid distance moduli from {\it HST}. We find no statistically significant difference between the two sets of distances measured between the same anchor and SN Ia hosts but with independent distance indicators.  We have not attempted to account for any further population-based differences intrinsic to the TRGB between the field of NGC 4258 and the mean of the SN Ia hosts, but believe these are limited due to the limited metallicity and age differences expected in \emph{F090W} and our restricted color window \citep{Anand_2024ApJ...966...89A}.

%%%%%%%%%%%%%%%%%%%%%%%%%%%%%%%%%%%%%%
\begin{deluxetable*}{cccccccccccccccc}
\tabletypesize{\tiny}
\label{tab:Distances}
\tablehead{
\colhead{Galaxy} &
\colhead{E(B-V)} &
\colhead{R} &
\colhead{NBT} &
\colhead{TRGB} &
\colhead{$\sigma$} &
\colhead{TRGB} &
\colhead{$\sigma$} &
\colhead{$\Delta$TRGB} &
\colhead{$\mu_0$(TRGB)} &
\colhead{$\sigma$} &
\colhead{$\mu_0$(TRGB)} &
\colhead{$\sigma$} &
\colhead{$\mu_0$} &
\colhead{$\sigma$} &
\colhead{$\sigma$ (var)}\\
\colhead{} &
\colhead{} &
\colhead{} &
\colhead{} &
\colhead{s=0.10} &
\colhead{s=0.10} &
\colhead{s=0.05} &
\colhead{s=0.05} &
\colhead{(s=0.10$-$s=0.05)} &
\colhead{s=0.10} &
\colhead{s=0.10} &
\colhead{s=0.05} &
\colhead{s=0.05} &
\colhead{Cepheid} &
\colhead{Cepheid} & 
\colhead{}}
\startdata
NGC$\,$1448 & 0.012 & 3.2 & 7392 & 26.97 & 0.02 & 27.02 & 0.05 & $-$0.05 & 31.34 & 0.05 & 31.39 & 0.07 & 31.30 & 0.05 & 0.02 \\
NGC$\,$1559 & 0.026 & 3.8 & 1582 & 27.13 & 0.01 & 27.14 & 0.01 & $-$0.01 & 31.50 & 0.05 & 31.51 & 0.05 & 31.50 & 0.07 & 0.01 \\
NGC$\,$2525 & 0.052 & 2.1 & 8225 & 27.50 & 0.03 & 27.44 & 0.07 & 0.06    & 31.87 & 0.06 & 31.81 & 0.09 & 32.06 & 0.11 & 0.04 \\
NGC$\,$3370 & 0.028 & 1.7 & 617 & 27.82 & 0.03 & 27.82 & 0.07 & 0       & 32.19 & 0.06 & 32.19 & 0.08 & 32.13 & 0.06 & 0.02 \\
NGC$\,$3447 & 0.026 & 1.9 & 678 & 27.56 & 0.04 & 27.55 & 0.08 & 0.01    & 31.93 & 0.06 & 31.92 & 0.09 & 31.95 & 0.05 & 0.01 \\
NGC$\,$5584 & 0.035 & 2.6 & 1582 & 27.58 & 0.06 & 27.43 & 0.10 & 0.15    & 31.95 & 0.08 & 31.8  & 0.11 & 31.77 & 0.06 & 0.07 \\
NGC$\,$5643 & 0.161 & 3.2 & 21032 & 26.20 & 0.01 & 26.21 & 0.01 & $-$0.01 & 30.57 & 0.06 & 30.58 & 0.06 & 30.55 & 0.06 & 0.01 \\
NGC$\,$5861 & 0.095 & 1.5 & 6199 &27.73 & 0.03 & 27.73 & 0.10 & 0       & 32.10 & 0.06 & 32.10 & 0.11 & 32.23 & 0.11 & 0.02 \\
\enddata
\caption{TRGB and Cepheid distances to NGC$\,$1448, NGC$\,$1559, NGC$\,$2525, NGC$\,$3370, NGC$\,$3447, NGC$\,$5584, and NGC$\,$5643. Cepheid distances are from \cite{Riess_2022ApJ...934L...7R}, fit variant 10. Distance errors do not include an 0.032 mag uncertainty from the maser distance for either Cepheids or TRGB.  TRGB distances include a (correlated across the host galaxies) measurement error for NGC$\,$4258 of 0.045 mag from \cite{Anand_2024ApJ...966...89A}. We adopt 15$\%$ of the extinction value as a systematic uncertainty unless it is less than 0.01~mag, in which case we instead adopt a full 1/2 of the extinction value as the uncertainty. We also add in quadrature a systematic uncertainty from measurement and peak choices with values and measurement variants listed in Table \ref{tab:Distances} and Table \ref{tab:Variants}, respectively. Columns from left to right: 1) Galaxy name, 2) E(B-V) from NED, 3) contrast ratio (R) defined from \cite{Wu_2023ApJ...954...87W}, 4) number of stars 1 magnitude fainter than the measured TRGB (NBT) with smoothing s = 0.05~mag,  5) TRGB measured here with a smoothing value of 0.10~mag, 6) error on that TRGB measurement, 7) TRGB measured here with a smoothing value of 0.05~mag, 8) error on that TRGB measurement, 9) difference between the TRGBs measured with s=0.05 and s=0.10~mag, 10) TRGB distance modulus measured here using a smoothing value of 0.10~mag, 11) error on that TRGB distance modulus, 12) TRGB distance modulus measured here using a smoothing value of 0.05~mag, 13) error on that TRGB distance modulus, 14) Cepheid distance \citep[fit variant 10; ][]{Riess_2022ApJ...934L...7R}, 15) error on the Cepheid distance, 16) dispersion of variants (measurement choices), may be considered as a systematic error on an individual distance measurement \ref{tab:Variants}.}
\end{deluxetable*}
%%%%%%%%%%%%%%%%%%%%%%%%%%%%%%%%%%%%%%

%%%%%%%%%%%%%%%%%%%%%%%%%%%%%%%%%%%%%%

\begin{figure}
\epsscale{1}
\plotone{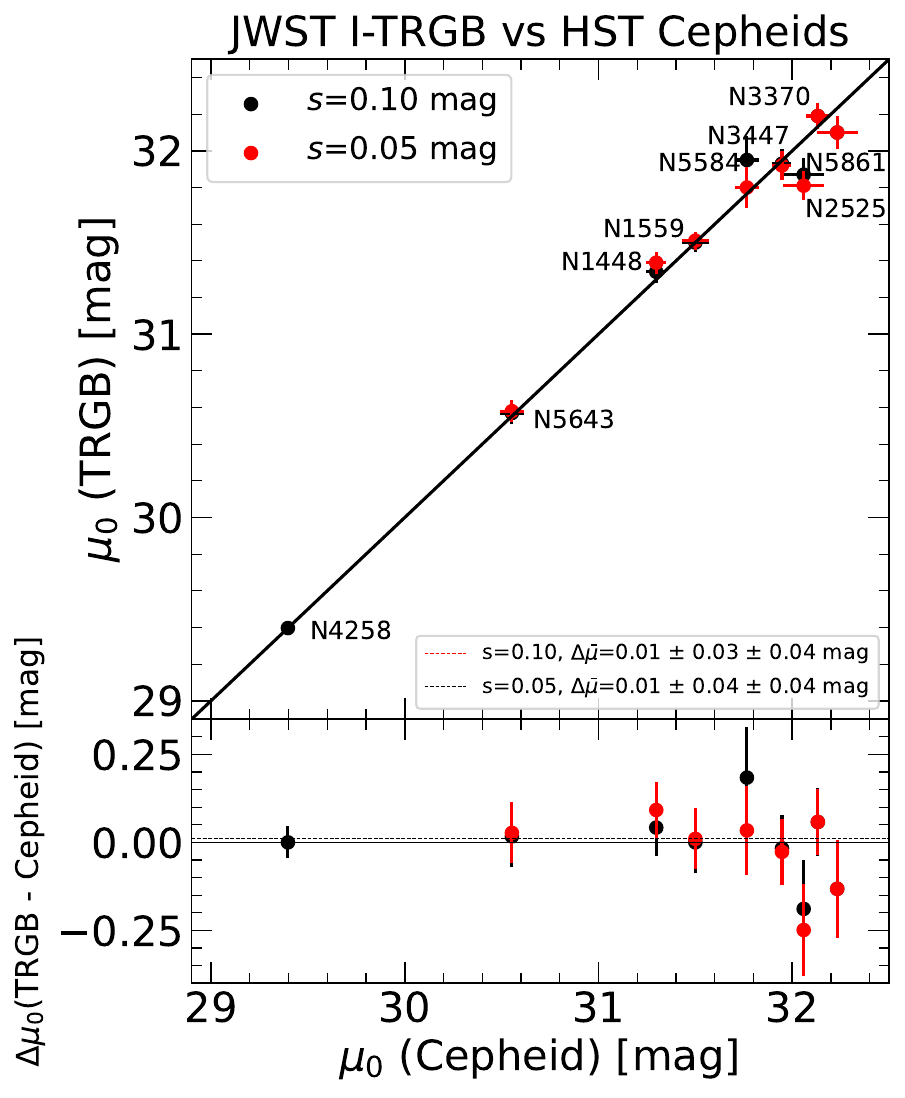} 
\caption{Comparison between the TRGB distances to NGC$\,$1448, NGC$\,$1559, NGC$\,$2525, NGC$\,$3370, NGC$\,$3447, NGC$\,$5584, NGC$\,$5643, and NGC $\,$5861 from this study and the Cepheid-based distances from \cite{Riess_2022ApJ...934L...7R} (fit variant 10). We compare to fit variant 10 from \citep{Riess_2022ApJ...934L...7R} for consistency of using only NGC$\,$4258 as a geometric reference. The top panel plots the TRGB distance moduli as a function of the Cepheid distance moduli. The bottom panel shows the residuals. The red and black points use TRGB measurements with 0.05 and 0.10~mag, respectively. The distance errors do not include a systematic error of 0.032~mag from the common anchor NGC$\,$4258 \citep{Reid_2019ApJ...886L..27R}. The weighted means shown include the variants error from Table \ref{tab:Distances} added in quadrature.}
\label{fig:distances}
\end{figure}
%%%%%%%%%%%%%%%%%%%%%%%%%%%%%%%%%%%%%%

The choice of algorithm and selection parameters can impact final TRGB measurements. One compelling approach is to optimize the measurement parameters to minimize scatter in the field-to-field variations (see \citealt{Wu_2023ApJ...954...87W, Scolnic_2023ApJ...954L..31S, Li_CATs_2023ApJ...956...32L}), however, the limited number of \emph{JWST} fields available to measure the TRGB within our host galaxies prevent us from performing such an optimization. To investigate the effects of how different parameter choices can impact our measurements, we explore 14 combinations of different luminosity function smoothing, color selection, and zero-point choice. We list the different measurement variants explored in this analysis, together with the weighted mean difference between the TRGB and {\it HST} Cepheid distances for each variant, in Table \ref{tab:Variants}. We include a thirteenth variant using the zero-point from \cite{Newmann_JWST_2024arXiv240603532N}, adjusted for the Vega$-$Vega photometric system which has differences of $\sim$0.04~mag in \emph{F090W} and $\sim$0.02~mag in \emph{F150W}. We also add a fourteenth variant where we use the median of all TRGB measurements (three smoothings, four color ranges) to compare with Cepheids.  We add the standard deviation of the measured TRGBs across the smoothing and color range variants for each galaxy, excluding the variants using the zero-point from \cite{Newmann_JWST_2024arXiv240603532N} and median TRGB, as a systematic error to our measurements, listed in Table \ref{tab:Distances}. We find that for all variants, the mean differences between the TRGB and Cepheid distances across all galaxies for a given variant are consistent with zero to within their statistical uncertainties with a full range of 0.00 to 0.02 mag.  

In Fig.~\ref{fig:distances}, the error in the mean $\mu_{TRGB}-\mu_{Cepheid}$ of 0.03$-$0.04 mag is a statistical uncertainty.  There is also a systematic uncertainty in the mean TRGB and Cepheid comparison from the measurement uncertainty of the TRGB in NGC$\,$4258 of $\pm 0.04$ mag \citep{Anand_2024ApJ...966...89A} which pertains to all calibrated TRGB distances and likewise for Cepheids in NGC$\,$4258 of $\pm 0.02$ mag, or $\pm 0.045$ mag when they are combined.  This systematic term due to the measurement of NGC$\,$4258 for both methods is the dominant term in the comparison and may only be reduced in the future with the use of additional observations of NGC$\,$4258.  The maser distance uncertainty of $\pm 0.032$ mag should not be considered in this distance comparison as it is common to both the Cepheid and TRGB distances.

We can also compare the distances to NGC$\,$1559 and NGC$\,$5584 from \cite{Anand_2024ApJ...966...89A}, who used the same dataset as here. \cite{Anand_2024ApJ...966...89A} found distances of $\mu_0$ = 31.49 $\pm$ 0.07~mag ($\Delta$ = 0.02 mag) and 31.80 $\pm$ 0.08~mag ($\Delta$ = 0.00 mag) to NGC$\,$1559 and NGC$\,$5584, respectively, when comparing their results to ours from the same smoothing scale (for NGC$\,$1559, we now have combined both visits to obtain higher S/N photometry). The TRGB distance for NGC$\,$1559 is also in agreement with that measured with Miras in \cite{Huang_2020ApJ...889....5H} of $\mu_0$ = 31.41 $\pm$ 0.050 $\pm$ 0.060~mag. 

An unavoidable feature of the TRGB method of edge detection is a potential ambiguity when multiple Sobel peaks of similar height are present.  For NGC$\,$5584 there are two Sobel peaks separated by 0.2$-$0.3 mag. Because the peaks are similar in height, different measurement choices may cause changes in which peak is higher. To account for this possibility, we provide an additional individual distance uncertainty listed in Table \ref{tab:Distances} estimated from the variation of individual results of variants in Table \ref{tab:Variants}. This uncertainty should be added for citing a single distance estimate for each host.  A similar situation involving two peaks at similar heights, but at a smaller level, 0.05 mag, is seen for NGC$\,$1448 and in the reverse direction for NGC$\,$2525. 

The Extragalactic Distance Database  \citep[EDD;][]{Anand_EDD_2021AJ....162...80A} and CCHP \citep{Freedman_Tensions_2021ApJ...919...16F} teams have measured \emph{HST} TRGB distances to NGC$\,$1448 and NGC$\,$5643, which can provide an interesting comparison with the \emph{JWST} distances found here. These distances also reflect differences in TRGB measurement methodology; the EDD team uses a model-based approach \citep{Makarov_2006AJ....132.2729M, Wu_2014AJ....148....7W}, while the CCHP team uses a Sobel filter applied on a Gaussian-windowed, Locally Weighted Scatterplot Smoothing (GLOESS) smoothed luminosity function \citep{Hatt_2017ApJ...845..146H}. We list this comparison in Table \ref{tab:N1559_N5643_Comparison}. For NGC$\,$1448, these teams independently found distance moduli of 31.38~mag (EDD) and 31.32~mag (CCHP), compared to 31.34~mag and 31.39~mag (different levels of smoothing) found here resulting in a standard deviation of 0.03~mag. For NGC$\,$5643, these team found 30.47~mag (EDD) and 30.48~mag (CCHP), compared to 30.57~mag and 30.58~mag (which agrees well with Cepheids) found here resulting in a standard deviation of 0.05~mag. 

%%%%%%%%%%%%%%%%%%%%%%%%%%%%%%%%%%%%%%
\begin{deluxetable*}{ccc}
\label{tab:N1559_N5643_Comparison}
\tablehead{\colhead{Galaxy} & \colhead{Reference} & \colhead{$\mu_0$(TRGB)}}
\startdata
NGC 1448 & \cite{Anand_EDD_2021AJ....162...80A} & 31.38  \\
NGC 1448 & \cite{Freedman_Tensions_2021ApJ...919...16F} & 31.32 \\
NGC 1448 & Here & 31.39 (s=0.05), 31.34 (s=0.10)   \\
NGC 5643 & \cite{Anand_EDD_2021AJ....162...80A} & 30.47  \\
NGC 5643 & \cite{Freedman_Tensions_2021ApJ...919...16F} & 30.48  \\
NGC 5643 & Here & 30.58 (s=0.05), 30.57 (s=0.10)  \\
\enddata
\caption{Summary table for distances to NGC 1448 and NGC 5643 from EDD \citep{Anand_EDD_2021AJ....162...80A}, CCHP \citep{Freedman_Tensions_2021ApJ...919...16F}, and here.}
\end{deluxetable*}

%%%%%%%%%%%%%%%%%%%%%%%%%%%%%%%%%%%%%%
\begin{deluxetable*}{ccccccc}
\label{tab:Variants}
\tablehead{\colhead{Smoothing [mag]} & \colhead{Color Cuts [mag]} & \colhead{$\bar{\mu}_{TRGB - Cepheid}$ [mag]} & \colhead{$\sigma$ [mag]} & Calibration & \colhead{$\chi^2$} & \colhead{Weighted Scatter [mag]}}
\startdata
0.05 & 1.15, 1.65 & 0.022 & 0.034 &\cite{Anand_2024ApJ...966...89A} & 1.103 & 0.10 \\
0.05 & 1.15, 1.75 & 0.018 & 0.034 &\cite{Anand_2024ApJ...966...89A} & 1.363 & 0.11 \\
\textbf{0.05} & \textbf{1.30}, \textbf{1.75} & \textbf{0.007} & \textbf{0.037} &\cite{Anand_2024ApJ...966...89A} & \textbf{0.712} & \textbf{0.09} \\
0.07 & 1.15, 1.65 & 0.014 & 0.033 &\cite{Anand_2024ApJ...966...89A} & 1.008 & 0.09 \\
0.07 & 1.15, 1.75 & 0.014 & 0.033 &\cite{Anand_2024ApJ...966...89A} & 1.317 & 0.11 \\
0.07 & 1.30, 1.75 & 0.013 & 0.034 &\cite{Anand_2024ApJ...966...89A} & 1.083 & 0.10 \\
0.10 & 1.15, 1.65 & 0.011 & 0.031 &\cite{Anand_2024ApJ...966...89A} & 1.057 & 0.09 \\
0.10 & 1.15, 1.75 & 0.015 & 0.031 &\cite{Anand_2024ApJ...966...89A} & 1.121 & 0.09 \\
0.10 & 1.30, 1.75 & 0.014 & 0.032 &\cite{Anand_2024ApJ...966...89A} & 0.961 & 0.09 \\
0.10 & 1.30, 1.75 & 0.005 & 0.032 &\cite{Newmann_JWST_2024arXiv240603532N} & 0.882 & 0.08 \\
0.15 & 1.15, 1.65 & 0.007 & 0.031 &\cite{Anand_2024ApJ...966...89A} & 1.139 & 0.09 \\
0.15 & 1.15, 1.75 & 0.014 & 0.030 &\cite{Anand_2024ApJ...966...89A} & 1.138 & 0.09 \\
0.15 & 1.30, 1.75 & 0.012 & 0.031 &\cite{Anand_2024ApJ...966...89A} & 1.015 & 0.09 \\
\hline 
All Smoothings & All Color Ranges & 0.008 & 0.031 &\cite{Anand_2024ApJ...966...89A}& 0.996 & 0.09 \\
\enddata
\caption{Summary table for the weighted mean differences between the TRGB and Cepheid distances (TRGB $-$ Cepheid) to NGC$\,$1448, NGC$\,$1559, NGC$\,$2525, NGC$\,$3370, NGC$\,$3447, NGC$\,$5584, NGC$\,$5643, and NGC$\,$5861 across various measurement variants. Cepheid distances are from \cite{Riess_2022ApJ...934L...7R}, fit variant 10. The error in the mean difference, TRGB$-$Cepheid, does not include a systematic uncertainty of 0.03 mag for the TRGB measurement error in NGC$\,$4258 from \cite{Anand_2024ApJ...966...89A}. Measurements use `simple' smoothing (i.e. no weighting, see \citealt{Wu_2023ApJ...954...87W}) and do not include the measurement choice error listed in Table \ref{tab:Distances} except for the last variant that uses the median TRGB across all smoothing and color range variants. Bolded values correspond to the baseline variant that uses $s=$~0.05~mag smoothing with a color range of 1.3 $<$ \emph{F090W - F150W} $<$ 1.75~mag. For the variant that uses the \cite{Newmann_JWST_2024arXiv240603532N}, we use the errors listed in their Tables 4 and 6 added in quadrature with the TRGB measurement and foreground extinction errors. Columns from left to right are: Smoothing value, color cuts, mean difference between TRGB and Cepheid distances, error on the mean difference, chi-squared, and weighted scatter.}
\end{deluxetable*}
%%%%%%%%%%%%%%%%%%%%%%%%%%%%%%%%%%%%%%
\newpage
\section{Discussion} \label{sec:Discussion}

We measure \emph{F090W} TRGB-based distances to NGC$\,$1448, NGC$\,$1559, NGC$\,$2525, NGC$\,$3370, NGC$\,$3447, NGC$\,$5584, NGC$\,$5643, and NGC$\,$5861 using \emph{JWST} NIRCam observations and test their consistency with \emph{HST} Cepheid-based distances to those same galaxies. We find excellent agreement between the two independent sets of distance measurements, which were measured between the same hosts and anchor but using data taken with different telescopes and distance indicators.

\begin{figure}
\epsscale{1}
\plotone{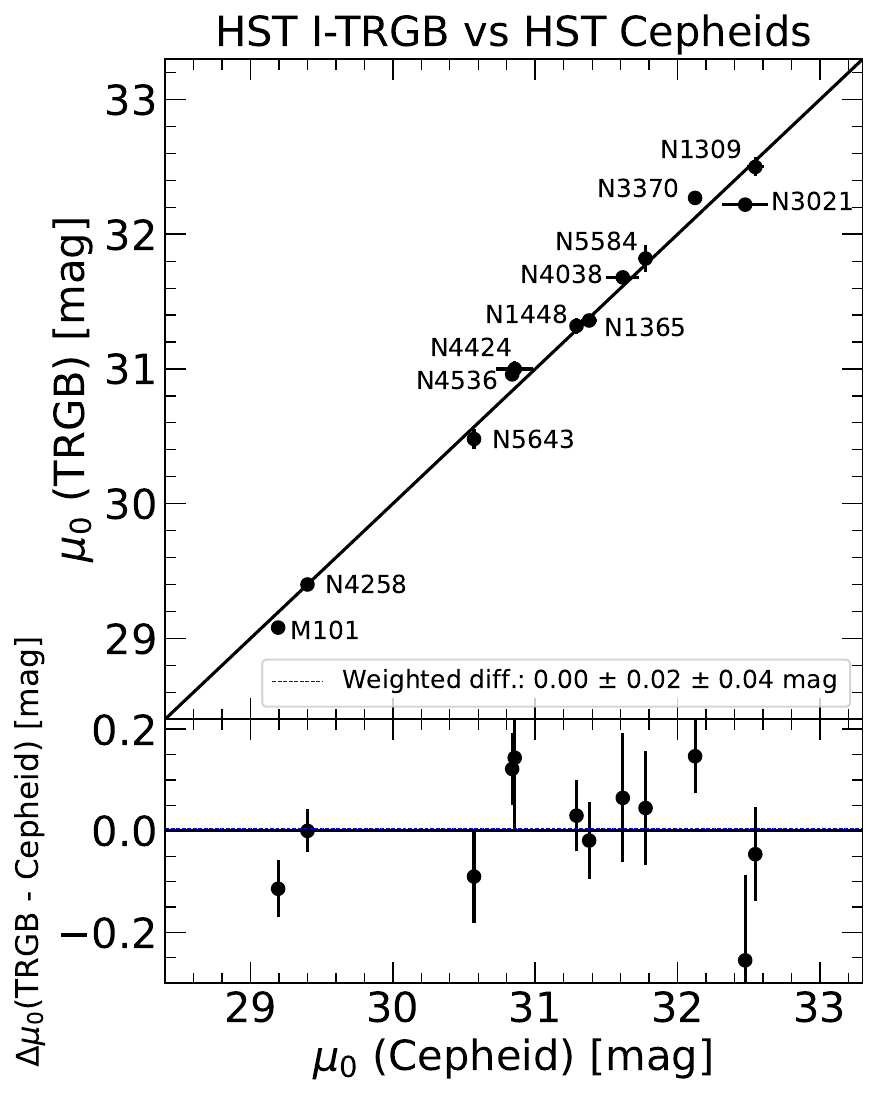} 
\caption{Comparison between the distances to M101, NGC$\,$5643, NGC$\,$4536, NGC$\,$4424, NGC$\,$1448, NGC$\,$1365, NGC$\,$4038, NGC$\,$5584, NGC$\,$3370, NGC$\,$3021, and NGC$\,$1309 based on TRGB \citep{Freedman_Tensions_2021ApJ...919...16F} and  Cepheids \citep{Riess_2022ApJ...934L...7R}. We add the distance to NGC$\,$4258 using the water maser from \cite{Reid_2019ApJ...886L..27R} for reference. The distance errors do not include a systematic error of 0.032~mag from the common anchor NGC$\,$4258 \citep{Reid_2019ApJ...886L..27R}.}
\label{fig:distances_HST}
\end{figure}

Examining the Hubble Tension necessitates close scrutiny of the consistency and systematics of distance indicators. Based on the measurements made in this study, we find no evidence of a bias or systematic difference between the Cepheid and TRGB methods that would cause a difference translatable to $H_0$ that can solve the Hubble Tension. This finding is in agreement with \cite{Freedman_2023JCAP...11..050F}, who find, ``the excellent agreement between the published Cepheid distances in Riess et al. (2022) and TRGB distances in Freedman et al. (2021), which in the mean, agree to 0.007 mag". While the recent analysis in \cite{Freedman_2024arXiv240806153F} found a 2.5$\%$ disagreement between TRGB and Cepheid distances, they do not include systematic measurement errors in NGC 4258 for the two standard candles of 0.04 and 0.09~mag, respectively. Sufficiently accounting for these terms in their comparison decreases the difference to below 1 $\sigma$ and provides no evidence of a significant systematic difference (see Appendix B in \cite{Riess_2024arXiv240811770R}. Notably, the primary paper describing the TRGB results in \cite{Freedman_2024arXiv240806153F} (Hoyt et al., in prep) is not yet available, preventing us from making any further comparisons.
In addition, Fig.~\ref{fig:distances_HST} shows a comparison between Cepheid and TRGB distances, both measured with \emph{HST}, using TRGB and Cepheid distances from \cite{Freedman_Tensions_2021ApJ...919...16F} and \cite{Riess_2022ApJ...934L...7R}, respectively. The TRGB and Cepheid distances in Fig.~\ref{fig:distances_HST} yield a mean difference of 0.00 $\pm$ 0.02 (stat) $\pm$ 0.04 (sys)~mag, also consistent with no difference even without additional consideration of populations differences in TRGB measurements \citep{Anderson_2024ApJ...963L..43A, Koblischke_2024arXiv240619375K}. This comparison will benefit from more \emph{JWST} TRGB measurements in the future (for instance, from GO-1995; \citealt{Freedman_2021jwst.prop.1995F}). As the Hubble Tension corresponds to 5$\mathrm{log_{10}}$(73/67.5) $\sim$ 0.17 mag, the demonstrated consistency is meaningful and is inconsistent with providing evidence for a solution to the Hubble Tension at 3$\sigma$.

%%%%%%%%%%%%%%%%%%%%%%%%%%%%%%%%%%%%%%

%%%%%%%%%%%%%%%%%%%%%%%%%%%%%%%%%%%%%%

We do not yet report a value of $H_0$ from the \emph{JWST}-calibrated SN subsamples because they are too small and the anchors too few to be competitive with the \emph{HST} sample of 42 SN Ia and 4 anchors.  The uncertainty in $H_0$ would be 2.5 times greater with $\sigma \sim$ 2-2.5 km/s/Mpc vs. 0.9  km/s/Mpc for \emph{HST}, which would trivially remove the Hubble Tension simply by inflating the error in $H_0$.  Rather, the greater power of these presently small \emph{JWST} samples comes from a direct comparison of what they can both measure in common, the distances between the same anchor and SN hosts.  

It will be important to also scrutinize the difference in direct $H_0$ measurements that arise from different treatments of SNe~Ia. \cite{Scolnic_2023ApJ...954L..31S} find that the addition of peculiar flow corrections and cross-calibration of datasets which are included in the Pantheon and Pantheon+ analyses increase $H_0$ measurements each by 0.5 and 1.1 km/s/Mpc from the \cite{Freedman_2021ApJ...919...16F} analysis, bringing it closer in line with the value obtained from Cepheids.  An under recognized source of variations in determinations of $H_0$ is also the make-up of the SN Ia calibration sample which can produce fluctuations of $\sim$ 1$-$2 km/s/Mpc due to small sample statistics.

Crosschecks between standard candles can be made more comprehensive with additional, independent standard candles such as Miras and carbon stars. The carbon star method, also called the J-region Asymptotic Giant Branch (JAGB), originated in the 1980s \citep{Richer_1981ApJ...243..744R, Richer_1984ApJ...287..138R, Richer_1985ApJ...298..240R, Pritchet_1987ApJ...323...79P, Cook_1986ApJ...305..634C} and has since been revived and further pioneered in the 2000s \citep{Battinelli_2005AA...442..159B, Ripoche_2020MNRAS.495.2858R, Madore_2020ApJ...899...66M, Freedman_2020ApJ...899...67F,  Parada_2021MNRAS.501..933P, Parada_2023MNRAS.522..195P, Zgirski_2021ApJ...916...19Z, 
Lee_MW_2021ApJ...923..157L,
Lee_WLM_2021ApJ...907..112L, Madore_MW_2022ApJ...938..125M,
Lee_M33_2022ApJ...933..201L, Lee_M31_2023_ApJ...956...15L, Lee_JWST_2024ApJ...961..132L, Li_2024ApJ...966...20L}. Although the JAGB method still requires further development and standardization (see discussions of asymmetric luminosity functions and metallicity effects in \citealt{Parada_2021MNRAS.501..933P, Parada_2023MNRAS.522..195P, Li_2024ApJ...966...20L}), the great luminosities of carbon stars in the near-infrared allow the JAGB to reach farther than the \emph{I}- and \emph{J}-band TRGB, potentially allowing for a more comprehensive crosscheck of Cepheid variables via galaxy distances.

The TRGB measurements made here can benefit from further analysis, such as a future contrast ratio calibration (see CATs; \citealt{Wu_2023ApJ...954...87W, Scolnic_2023ApJ...954L..31S, Li_CATs_2023ApJ...956...32L}) or via modelling of the luminosity function \citep{Makarov_2006AJ....132.2729M, Anand_Local_Group_2019ApJ...880...52A, Anand_2024ApJ...966...89A} instead of edge-detection algorithms (Anand et al., in prep). However, even without these additions, the consistency of the TRGB-based distance measurements with those of Cepheid-based distances measurement do not show evidence of a bias or systematic offset in Cepheid distances that would resolve the Hubble Tension. Similar analyses in the future will also benefit from further studies investigating the effects of red giant diversity to improve the accuracy of TRGB distances \citep[such as in][]{Anderson_2024ApJ...963L..43A, Koblischke_2024arXiv240619375K}. Future observations of TRGB fields with \emph{HST} (for instance, from GO-17520;  \citealt{Breuval_2023hst..prop17520B}), \emph{JWST}, and the \emph{Roman} Space Telescope will provide more opportunities to scrutinize the second rung of the distance ladder by providing a more extensive set of observations that can be used to compile a more comprehensive comparison between these distance indicators. Specifically, wide-field \emph{JWST} observations within the anchor galaxy NGC~4258 will help reduce systematic uncertainties with regards to the absolute calibration of the TRGB, and deep \emph{JWST} observations of galaxies with the highest levels of mismatch between \emph{HST} Cepheid and \emph{HST} TRGB distances (e.g. NGC~3021, NGC~4038/9; \citealt{Jang_2017ApJ...836...74J}) will help elucidate any individual discrepancies between the two techniques. Lastly, a fully independent distance ladder from the traditional Cepheid+SNe~Ia route is also under construction, via the usage of the TRGB combined with Surface Brightness Fluctuations as the final rung \citep{Anand_Fornax_2024arXiv240503743A}.

\begin{acknowledgments}

We are indebted to all of those who spent years and even decades bringing {\it JWST} to fruition. We thank Yukei Murakami for helpful conversations. SL is supported by the National Science Foundation Graduate Research Fellowship Program under grant number DGE2139757. GSA acknowledges financial support from JWST GO-1685 and JWST GO-2875. DS is supported by Department of Energy grant DE-SC0010007, the David and Lucile Packard Foundation, the Templeton Foundation and Sloan Foundation.  RLB is supported by the National Science Foundation through grant number AST-2108616. RIA is funded by the SNSF through an Eccellenza Professorial Fellowship, grant number PCEFP2\_194638.

This research made use of the NASA Astrophysics Data System. This work is based on observations made with the NASA/ESA/CSA JWST. The data were obtained from the Mikulski Archive for Space Telescopes at the Space Telescope Science Institute, which is operated by the Association of Universities for Research in Astronomy, Inc., under NASA contract NAS 5-03127 for JWST. These observations are associated with programs $\#$1685  and $\#$2875. The JWST data presented in this article were obtained from the Mikulski Archive for Space Telescopes (MAST) at the Space Telescope Science Institute. The specific observations analyzed can be accessed via \dataset[DOI: 10.17909/556t-7522].

The Digitized Sky Surveys were produced at the Space Telescope Science Institute under U.S. Government grant NAG W-2166. The images of these surveys are based on photographic data obtained using the Oschin Schmidt Telescope on Palomar Mountain and the UK Schmidt Telescope. The plates were processed into the present compressed digital form with the permission of these institutions.

\end{acknowledgments}

\vspace{5mm}
\facilities{JWST (NIRCam)}

\software{astropy \citep{astropy_2013A&A...558A..33A,astropy_2018AJ....156..123A, astropy_2022ApJ...935..167A}, 
SAOImage \citep{SAO_Image_2003ASPC..295..489J}, 
DOLPHOT \citep{Dolphin_2000PASP..112.1383D, Dolphin_2016ascl.soft08013D, Weisz_2024ApJS..271...47W}}

\bibliography{sample631}{}
\bibliographystyle{aasjournal}

\end{document}